\documentclass[12pt]{article}

\usepackage[margin = 2cm]{geometry}

\usepackage{graphicx}
\usepackage{amssymb}

\usepackage{lineno}
\usepackage[title]{appendix}

\usepackage{multirow}
\usepackage{setspace}
\usepackage{natbib}
\usepackage{booktabs}
\usepackage{caption}
\usepackage{amsmath}
\usepackage{hyperref}
\usepackage{ccicons}
\usepackage[T1]{fontenc}
\usepackage{enumitem}
\usepackage[table,xcdraw]{xcolor}
\usepackage{floatrow}
\usepackage{subcaption}
\usepackage{authblk}
\usepackage{booktabs}
\usepackage[normalem]{ulem}
\useunder{\uline}{\ul}{}
\graphicspath{ {././} }
\begin{document}


\title{A transformer-based model for default prediction in mid-cap corporate markets \footnote{\scriptsize NOTICE: This is a preprint of a published work. Changes resulting from the publishing process, such as editing, corrections,structural formatting, and other quality control mechanisms may not be reflected in this version of the document. Please cite this work as follows: Korangi, K., Mues, C. and Bravo, C. (2022). A transformer-based model for default prediction in mid-cap corporate markets. European Journal of Operational Research, DOI: https://doi.org/10.1016/j.ejor.2022.10.032. This work is made available under a \href{https://creativecommons.org/licenses/by/4.0/}{Creative Commons BY license}. \ccby}}

\author[1,2]{Kamesh Korangi}
\author[1,2]{Christophe Mues}
\author[3]{Cristi\'{a}n Bravo}

\affil[1]{Department of Decision Analytics and Risk, Southampton Business School, University of Southampton, University Road, SO17 1BJ, United Kingdom.}

\affil[2]{Centre for Operational Research, Management Sciences and Information Systems (CORMSIS), University of Southampton, University Road, SO17 1BJ, United Kingdom.}

\affil[3]{Department of Statistical and Actuarial Sciences, The University of Western Ontario,1151 Richmond Street, London, Ontario, N6A 5B7, Canada.}

\date{}

\maketitle
\begin{abstract}
In this paper, we study mid-cap companies, i.e.\@ publicly traded companies with less than US\$10 billion in market capitalisation. Using a large dataset of US mid-cap companies observed over 30 years, we look to predict the default probability term structure over the short to medium term and understand which data sources (i.e.\@ fundamental, market or pricing data) contribute most to the default risk. Whereas existing methods typically require that data from different time periods are first aggregated and turned into cross-sectional features, we frame the problem as a multi-label panel data classification problem.
To tackle it, we then employ transformer models, a state-of-the-art deep learning model emanating from the natural language processing domain.
To make this approach suitable to the given credit risk setting, we use a loss function for multi-label classification, to deal with the term structure, and propose a multi-channel architecture with differential training that allows the model to use all input data efficiently.
Our results show that the proposed deep learning architecture produces superior performance, resulting in a sizeable improvement in AUC (Area Under the receiver operating characteristic Curve) over traditional models. In order to interpret the model, we also demonstrate how to produce an importance ranking for the different data sources and their temporal relationships, using a Shapley approach for feature groups.
\end{abstract}


\section{Introduction} \label{sec:Introduction}
\par

Traditional credit risk models cater to individual consumers with empirical models (built by applying statistical or machine learning methods to large datasets). In contrast, corporate credit risk models are often theory-driven or may include a qualitative component. Rating agencies play an important role in determining corporate credit risk. That rating process is costly, and it also has a strong subjective component \citep{Frost2007,RonaTas2010}, which is often needed because, unlike with consumer credit risk models, the small number of firms may affect the quality of statistical models. The subjective component typically involves experts deciding how many notches to downgrade or upgrade a model-generated rating based on market conditions, recent events, and other criteria they seek to take into account. Although this approach is appropriate for large companies, such a qualitative assessment would not be scalable for the much larger population of small to medium-sized companies. Neither could we reapply the same quantitative approaches developed for consumer credit risk, as the default signal in the corporate setting comes from a complex combination of internal and external market conditions. In our work, we seek to remove the subjective component of the rating process by incorporating the additional data sources (such as market conditions), that would have previously required further qualitative assessment, directly into our quantitative deep learning models. As more data is collected over time, covering a wider range of market circumstances, the relationships learnt by these models will also evolve.  As a result, rating decisions made based on these models will be further fine-tuned as well. We use three data sources, i.e.\@ accounting data, pricing data and general market data, as part of a multi-channel deep learning model that predicts the default risk of mid-cap companies that are active in debt (bond or loan) markets.

Mid-cap firms (in short \lq{mid-caps}\rq) are defined in the US as firms with USD 1 to 10 billion market capitalisation and are possible constituents of the Dow Jones Wilshire Mid-cap index or S\&P 400 Mid-cap index. Their debt has a relatively short legal maturity period of around 5 to 10 years (compared to over 20 years for large-caps). The effective maturity of the debt can be as short as half the legal maturity, after considering embedded options and coupon rates that tend to be higher than those for large-caps. Mid-caps also tend to differ from large caps in terms of the relative credit risk they pose. In corporate debt markets, the mid-caps typically hold a non-investment grade credit rating, implying higher credit risk. Given that the listed mid-cap companies provide public data about their financial accounts, stock exchanges publish stock prices, and default history is available, lenders have all the data required to construct sophisticated credit risk models for them.
 	
Despite the availability of such data, building these models presents several challenges. First, the credit spreads or prices implied by the models often differ from what is empirically observed --- termed the `credit spread puzzle' by \citet{Amato2003}. This means that mid-cap credit risk is not accurately priced, which can lead to underestimation of potential losses. A second challenge is the difficulty in separating credit risk and market risk for mid-cap firms \citep{Jarrow2000}. Finally, the covenants in debt offerings and embedded options make the maturity and capital structure dependent on market conditions \citep{Liu2016}. All these issues make it difficult for lenders or investors to assess risk on a large scale, thus limiting access to credit for the companies involved. To address this, governments have established supporting institutions providing financing to mid-caps and small and medium-sized enterprises, such as the European Investment Bank (EIB) in Europe and the British Business Bank in the UK.

Another challenge in building corporate default prediction models lies in the time horizon of the prediction models. Most credit risk models study the probability of default over a one-year time horizon, due to business practices and regulatory frameworks such as the Basel Accords \citep{Supervision2003}. However, the time from financial distress to an actual default could easily be longer.  In the capital requirement models cited above, this is reflected by the maturity component of debt, but this is not usually captured by the probability of default (PD) models themselves. Several methods have been proposed to extend the models to longer horizons \citep{Duffie2007,Jardin2015,Altman2020}. Still, multi-horizon models are not widely implemented due to the lack of sufficient historical data under different macroeconomic conditions, changes in distribution of the variables, relationship drift between explanatory variables over time and changes in relationship with the dependent variable \citep{Jardin2012}. Instead, different models tend to be developed for different respective time horizons, and generally, ensemble models are used for better performance, making the modelling complex. We are interested in predicting the probability of default from a short-term horizon of several months to a medium-term horizon of one to three years, using a unified model. This is close to the effective maturity of these instruments and considers most lenders' investment horizons in this area of the market. 

The techniques used for default prediction modelling have evolved over time and remain an active area of research \citep{Dastile2020}. Traditionally, popular linear models such as the logit model or discriminant analysis require making a large number of discretionary decisions when handcrafting a set of predictive features (such as the choice of lookback period and aggregation functions), as well as making some restrictive assumptions about the distribution of the data or the functional form of the relationship between those features and default risk (such as linearity). 
In addition, large datasets may also require further feature selection \citep{Jones2017}. On the other hand, machine learning models allow for a large set of features and can handle non-linear relationships, which can produce predictive performance gains over linear models. However, integrating these different kinds of (often diverse) data sources remains challenging as the process to represent data becomes complex \citep{Mai2019}. Such data could include non-structured data (such as text or audio) and may contain a mix of high-frequency
and low-frequency data. 
Deep learning models \citep{LeCun2015} can cope not only with large amounts of data, but, using techniques such as multimodal learning, they can also handle different types of data effectively \citep{Ngiam2011}. Furthermore, they are able to identify non-linear correlations over longer time frames, which other methods could overlook. These properties make deep learning a promising approach for the mid-cap default prediction setting, as they allow us to use different forms of data alongside each other and capture how they affect default risk without the need for manual feature creation. 

Within the deep learning community, a variety of model types have emerged. Among these, the transformer-based deep learning models have recently produced state-of-the-art results in tasks involving other sequential data such as text, audio and video data \citep{Vaswani2017}. Unlike earlier deep learning models, transformers do not incorporate the position of a data point in a time series as relevant, which is a different design compared to Long Short-Term Memory (LSTM)-based models. Those employ recurrence as a key feature, using the present input and selected past information to arrive at a prediction. Instead, transformers use the whole past information along with the present to produce their predictions. LSTM models, originally developed by \citet{Hochreiter1997a}, have for some time been the common method of choice for time series or sequential data. Therefore, we also include LSTMs in our study as a benchmark against which to compare our proposed transformer-based architecture.

Although deep learning can help increase the accuracy of model predictions, interpreting how these predictions are derived presents an added challenge. We address this issue in two ways. Firstly, transformer models allow us to visually interpret the temporal relationships extracted from the data using attention heat-maps. In this paper, however, we put forward a second method and apply a Shapley approach \citep{Shapley1953} to quantify the relative importance of groups of variables and different lookback periods. This will allow us to gain deeper insights into the mid-cap risk structure.
 
Therefore, the three key research questions addressed in the paper are: 

\begin{enumerate}
\item  Can an effective transformer-based model be developed that uses accounting, pricing and market data for mid-cap default prediction? 

\item Can this architecture accurately predict a term structure for the probability of default over a short to medium-term horizon (3 months to 3 years)?

\item Which data sources and past time periods contribute most to the default risk estimates?
\end{enumerate}

The remainder of the paper is organised as follows. Section~\ref{sec:Literature Review} presents a literature review on corporate default risk modelling, discussing the popular models, studies on specific mid-cap issues and relevant machine learning research. Section~\ref{sec:Data Description} describes the data used in the paper. The proposed models and the benchmark models against which they are compared are described in Section~\ref{sec:Models}. Section~\ref{sec:ExperimentalSettings} discusses the experimental design, performance metrics, the Shapley group method and hyper-parameter tuning strategies.  Section~\ref{sec:Results} presents the results and highlights some discussion points relevant to mid-cap companies. Finally, Section~\ref{Conclusion} summarises the contributions and suggests future work.

\section {Literature review} \label{sec:Literature Review}
Corporate default prediction research has thus far focused on three types of approaches. All of these have also seen commercial implementations by rating agencies such as Standard \& Poor's (S\&P), Moody's and Fitch Ratings. 

The first approach is to build statistical models for default prediction, using accounting information from financial statements and applying econometric techniques. These models initially used univariate analysis \citep{Beaver1966}, later multivariate analysis \citep{Altman1968}, and they continue to be developed to the present day \citep{Altman2020}. S\&P and Fitch use this approach commercially and augment the models with expert opinions and industry-specific metrics. There are, however, limitations to these models. Accounting information could be restated by management and discretionary changes limit the predictive power of these models when companies are under financial stress \citep{Beaver2012}. 

The second set of models are structural models, which use a combination of accounting and pricing information, within an option-theoretic framework. \citet{Merton1974} developed the first such model using Black-Scholes option theory. Structural models are also used in commercial applications such as Moody's KMV model \citep{Crosbie2003}. Despite their ability to use current market price information to predict default, there are some limitations to these models as well. Assumptions on asset volatility need to be made as they are not observable and the firm capital structure needs to be simplified to quantify the value of debt as an option on the firm value. Also, default of the firm is endogenous to the model and occurs when the asset value drops below  outstanding debt \citep{Jarrow2000}. 

The third type of models are reduced-form models. They use mainly market information and especially credit spread information of public companies, applying arbitrage-free valuation techniques. \citet{Jarrow1995} were the first to propose such models in which both the  interest rate term structure and credit spread term structure are stochastic, unlike in previous models that assumed interest rates as fixed. Their main use has been in the pricing of credit derivatives of large firms. However, as they rely on public trading information and bond prices, they cannot be applied to private companies or companies with illiquid trading patterns or non-tradeable debt, which makes them unsuitable for mid-cap companies. 

Mid-cap companies present their own specific challenges to any of these credit risk models. \citet{Amato2003} first reported the phenomenon of the credit spread puzzle; i.e., they found that the difference between the model-based credit risk estimates and the empirical risk increases as credit ratings drop below investment grade, which is where most mid-cap companies are rated. \citet{DeJong2012} and \citet{Lin2011} have suggested the existence of a liquidity premium as one possible factor impacting the credit risk estimates for these companies. \citet{Beckworth2010} found monetary policy shocks to be another factor determining credit spreads, together with economic conditions. \citet{Acharya2013} further explain the puzzle by adding shocks to economic conditions through liquidity, especially for mid-cap companies with non-investment grade ratings. Later studies by \citet{Feldhutter2018} found the credit spread puzzle to be more pronounced for high yield or mid-cap companies, while large firms were less affected. \citet{Du2019} reduced the difference between model and empirical credit spreads by further improving the structural models, including uncertainty from asset risk. \citet{Bai2020} reject the existence of the credit spread puzzle, but their report uses credit default swap spreads, which is a different market to the bond market used in previous research. The latter is more relevant to mid-cap firms as they are much more dependent for their capital on bond and loan markets, compared to equity markets.

The second set of challenges that complicate mid-cap credit risk modelling arises from market risk factors. For any firm whose debt is traded, credit risk is not easily separable from market risk. This holds even more for mid-cap companies, whose debt is more correlated with equity indices than with treasury rates \citep{Jarrow2000}. Credit risk models hence need to incorporate a number of market-related factors and condense that information to an effective market representation which can be used to help determine probability of default. 

As the aforementioned studies show, modelling mid-cap credit risk is complex and different approaches consider a variety of factors. In this paper, we aim to bring together some of these strands by looking at accounting factors, general market factors and firm equity performance to estimate the probability of default or credit risk. We propose to tackle this problem with deep learning models and make a case for why they are more suitable for this task. 

In default or bankruptcy prediction, \citet{Tam1992} were one of the first to use (shallow) neural networks, showing they gave better performance compared to linear models such as those built using logistic regression. Also, \citet{Zhang1999} demonstrated that neural networks are sufficiently robust to deal with unseen data. \citet{Kim2010} applied Support Vector Machines (SVMs) to small and medium scale enterprise default prediction and reported greater accuracy. Later research continued with ensembles of model predictions. \citet{Alaka2018} reviewed different predictive models such as multi-layer neural networks, support vector machines, rough sets, case-based reasoning, decision trees, genetic algorithms, logistic regression and discriminant analysis models in the domain of bankruptcy or default prediction. They found that an ensemble of these models performed better but they noted that combining all of these models into a hybrid model needs informed study of the individual models. Apart from ensembles of classifiers, a recent meta-analysis of the literature by \citet{Dastile2020} identified another class of techniques that showed promising results --- deep learning models.

Compared to the former machine learning techniques, the number of papers in the area of credit risk modelling that have applied deep learning models, such as LSTMs, convolutional neural networks, and, most recently, transformers, is much smaller but growing. \citet{Kim2021} applied LSTM models to bankruptcy prediction for listed US firms between 2007-2019, and found that LSTM and ensemble models outperformed other techniques in accurately predicting bankruptcies. Since LSTMs are commonly used in other domains as well, we have, therefore, included them as one of the benchmark models in our work. \citet{Mai2019} applied convolutional neural networks to a large dataset containing textual data (from the 10-K reports on financial performance and risks submitted by company management), along with other accounting data, and found that deep learning models performed better. \citet{Stevenson2021} applied BERT (Bidirectional Encoder Representations from Transformers) to predict default in micro enterprises. They found textual data provided by a loan expert to be predictive of default but that similar information could also be captured by structured data.
 
Our approach differs from the above works by considering panel data and employing the encoder part of a transformer model to analyse such data (as opposed to the textual data to which they are more often applied).  \citet{Vaswani2017} first developed the transformer model, which introduces a multi-headed self-attention mechanism. This mechanism eschews recurrence so that the whole data input can be used. Also, it allows interactions between inputs when extracting relationships. Multiple heads also allow different relationships to be learnt. Transformer-based models have since significantly outperformed LSTM-based models in natural language tasks \citep{Lakew2018} and speech-related problems \citep{Karita2019}. Furthermore, \citet{Wen2022} reviewed recent developments in applying transformers to time series modelling, where they achieved similar levels of performance in multivariate time series forecasting and classification tasks. In our problem setting, the data input is panel data, which is commonly encountered in the credit risk literature, both in the consumer risk domain \citep{Leow2016} or for corporate risk \citep{Mai2019,Delis2020}. By applying transformers to such data, this study fills a gap in the literature.

Hence, the first contribution of our paper is that we are the first to propose a transformer-encoder model architecture to accurately estimate corporate default risk. Second, we select a framework for multimodal learning that can combine the different data sources and allows for a differential training approach where we can train each model separately.
 
Compared to traditional linear models, machine learning models tend to improve predictions but can come at the cost of reduced interpretability, which hinders their application in highly regulated areas such as credit risk \citep{Alaka2018}. Transformer models, even though they are complex, are arguably more interpretable than some other deep learning methods. For example, \citet{Wiegreffe2019} studied the attention weights after training the model and found them useful for explaining the model's predictions. Although these weights can help us understand the impact of certain variables, we are also interested in understanding the relative importance of each of the three data channels. For that purpose, we adopt
a suitable method based on Shapley values. Several such methods have been proposed to interpret a model \citep{Lundberg2017}, but as we aim to quantify the importance of a group of (rather than individual) variables, we follow the approach by \cite{Nandlall2019}. In so doing, we are able to make a third contribution, which is to answer questions about the relative importance of different data sources and study how the strength of these relationships varies over time.

Our fourth and final contribution is to the credit risk literature, as we show how multi-horizon probability of default estimates can be produced using a single deep learning model, and how this model produces good results not just in the short term but over a medium term of up to three years.

To benchmark the predictive performance of our proposed transformer model against other methods, we consider a series of methods including logistic regression, shallow neural networks, machine learning classifiers such as XGBoost, and other deep learning alternatives such as LSTMs and Temporal Convolutional Networks (TCN). XGBoost, a scalable decision tree-based ensemble learning algorithm developed by \citet{Chen2016} has achieved state-of-the-art results in many machine learning competitions, especially in classification tasks using structured data. The same technique applied to bankruptcy prediction also produced good results \citep{Zieba2016}. Second, Temporal Convolutional Networks (TCN) are another deep learning model that combines a series of techniques used in both sequence and image processing models. TCNs have been successfully used to classify time series data in health \citep{Sun2015, Lea2017} and other domains \citep{Pelletier2019}. We use the version of TCN developed by \citet{Bai2018} --- a generic architecture that can be applied to our task (see Section \ref{subsec:TCN}). Similarly to transformer models, TCNs have not yet been applied to default prediction in consumer or corporate credit risk either, as far as we are aware. Hence, by comparing our proposed transformer model to several powerful benchmark models, we add the necessary robustness to the findings of our study.

\section {Data} \label{sec:Data Description}

\begin{figure}[t]%[!ht]
\includegraphics[scale=0.55]{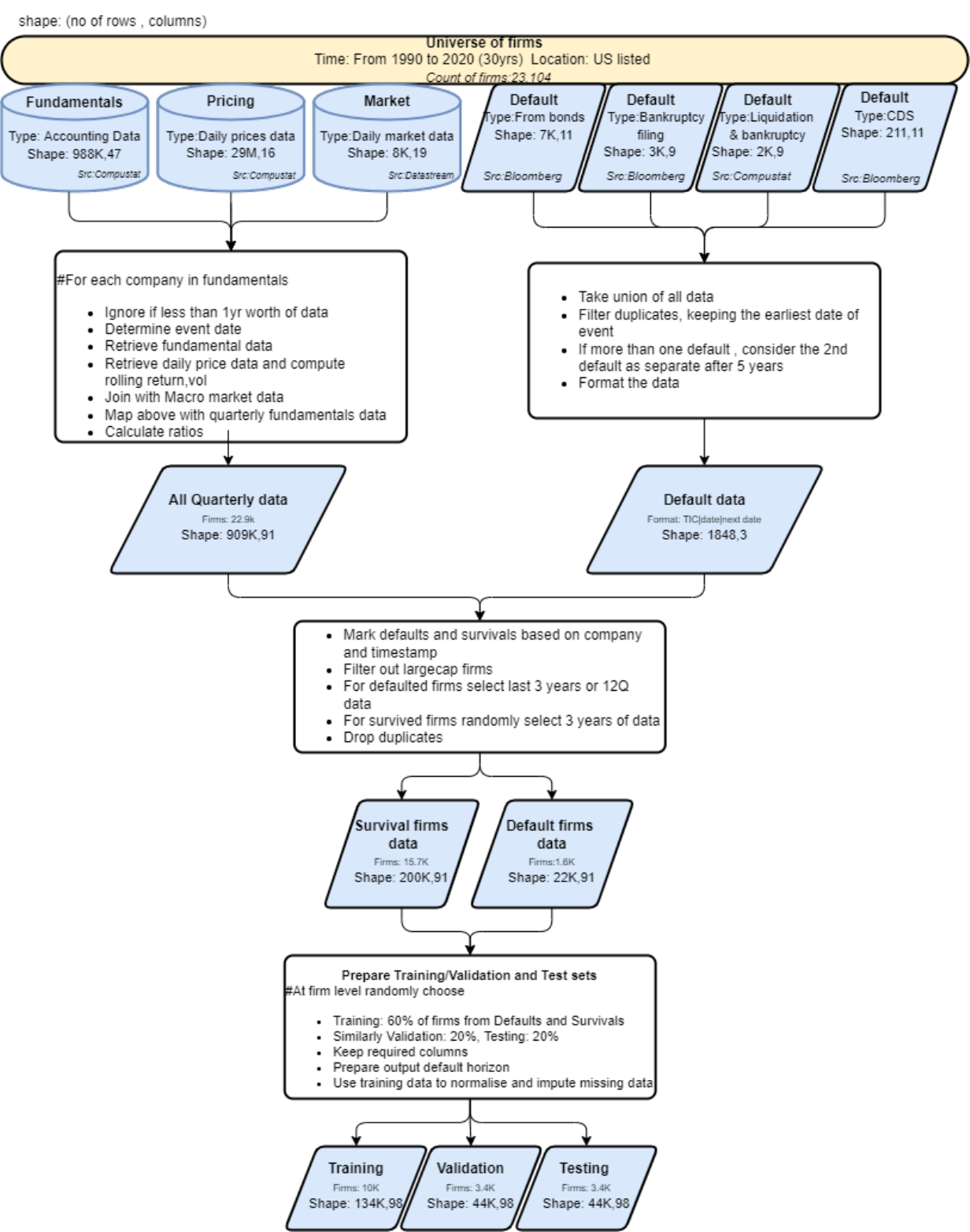}
\centering
  \caption{Data processing}
   \label{fig:data processing}
\end{figure}

We collected 30 years of data related to mid-cap companies listed in the US from 1990 to 2020, from the following sources: CRSP/Compustat for accounting data and pricing data, Bloomberg and CRSP for default information, and Datastream for market performance data. We exclude financial firms as their leverage and accounting measures are different from non-financial firms, following the standard practice in the corporate default prediction literature \citep{Shumway2001}. To be included in the sample, mid-cap companies are required to have a minimum of three years of financial accounting history. For details on how the data is processed, we refer to Figure~\ref{fig:data processing}.

\subsection{Data channels}
We distinguish between three different data sources (channels): 
 
\begin{enumerate}[label=(\roman*)]
\item Fundamental channel: This provides quarterly accounting data expressed as ratios observed at different time points. Sampling is done quarterly instead of over yearly intervals, as the latter would miss the accounting periods' seasonal volatility. The quarterly data is annualised using the last twelve months' metric such that all data is comparable. This data source is useful in capturing the firm's state at a specific time period or understanding how changes in those ratios may affect default risk.
We refer to \cite{Mai2019} for the financial ratios included.

\item Market channel: Quarterly market performance is collected over the same time period as the fundamental channel data. This data captures general market conditions and includes any financial ratios derived by combining accounting and market data. We included: the S\&P 500 Composite Price Index; 10-year US Treasury yield index; All corporate bonds (ICE Bank of America Corporate Index) and High yield bond index (ICE BofA US High Yield index).  

\item Pricing channel: Daily high, low and close history of each firm's equity prices. It consists of very few features, but they are collected at a much higher frequency than the other two channels, providing a detailed record of each firm's recent market valuation history.

\end{enumerate}
    
\subsection{Default definition and reporting event dates}\label{subsec:Default definition}
We define that a firm is in default if any one of the following criteria is satisfied:
the firm filed for bankruptcy; the company is under liquidation; a credit event has been declared as defined by the International Swaps and Derivatives Association (ISDA) which led to the triggering of Credit Default Swaps (CDS); or the firm has failed to pay interest or principal on any of its debt instruments. 

This is a broader default definition than simply identifying default on the basis of a bankruptcy filing. It is intended to capture most default scenarios at the earliest opportunity.  For example, failure to pay interest or principal is an early indicator of default, which predates a subsequent bankruptcy filing (if any).
CDS events also sometimes capture defaults earlier, as the market participants independently determine them. A CDS trigger might not push a company towards bankruptcy, but it could mean losses to its debt holders. This definition makes the predictive modelling more challenging as the firm's financial data might not yet have deteriorated to the same extent as with the traditional bankruptcy or liquidation filing approach. This is a more useful approach as it replicates the real-world scenario.

The timestamp that we record for each reporting event is also important to note. Here we take a different approach to the literature, by using the actual reporting date on which the financial results are published, which may differ from company to company. This approach avoids having to add an extra lag to the financial information as it is typically done.

\subsection{Target vector and data structure}
To be able to predict default over a short to medium-term horizon, we create a multi-label target vector, $Y$, consisting of binary variables, of the form \[ Y=[default_{3m}, default_{6m}, default_{9m}, default_{1y}, default_{2y},default_{3y}], \] with 1 denoting a default event, or 0 otherwise. We code all subsequent periods as a default from when the default occurred. For example, if default occurred 10 months after the timestamp, the vector would hold the values $[0, 0, 0, 1, 1, 1]$. This creates an incremental multi-label classification problem, where, as the time horizon increases, the class imbalance decreases, but the event becomes harder to predict. 

The observed inputs, $X$, for each firm are a matrix of dimensions $w\times f$, where $w$ is the maximum number of historical time periods and $f$ denotes the number of features (input variables) in the data. The input variables collected from the three channels are further preprocessed using standardisation and by treating outliers and missing data. We normalise the data using median and interquartile values and winsorise the data for values beyond six times the interquartile ranges. This limits the impact of severe outliers on the model parameters. We replace missing values with the median and add dummies to mark those replacements, since reporting gaps more frequently occur when firms are under financial stress and, thus, these data might not be missing at random.

\section{Models}\label{sec:Models}

In this section, we describe our novel Transformer Encoder model for Panel data (TEP), as well as another recent deep learning model against which it will be benchmarked, i.e.\@ Temporal Convolutional Networks (TCN). We omit describing our other models for brevity. For further information on logistic regression, shallow neural networks and XGBoost models, we refer the reader to \citet{Hastie2001}; LSTM models are explained in \citet{Goodfellow2016}.

\subsection{Transformer Encoder for Panel-data classification (TEP)}

Transformers have thus far been used mainly in the field of Natural Language Processing (NLP). These models incorporate a \emph{self-attention mechanism} to store learnt patterns. When looking at sequential data, this mechanism ensures that each data point is related to every other data point in the sequence. The architecture further allows for multiple attention heads, each of which can focus on a different aspect of the input, thereby extracting complex non-linear relationships. This ability makes Transformers different in how they handle sequence data. Unlike earlier sequence models based on Recurrent Neural Networks (RNNs), such as LSTMs, transformers take the whole sequence as an input and focus on multiple disjoint sequences to generate patterns. Whereas with NLP tasks where, depending on the document word count, the input sequence can be rather long, the sequence here is restricted to panel data, which is shorter and so does not face the same memory or computation costs. The standard model consists of an encoder and a decoder, as is typical in sequence-to-sequence models. During training, the encoder takes the numerical input and each of its heads learns different input aspects, thus creating a higher-order representation. The encoder output is transferred to the decoder. The decoder applies a similar self-attention mechanism to the output sequence to generate a representation and another self-attention layer to combine the encoder representation and earlier output sequence representation. This is passed through a dense feed-forward network to produce the final representation.
   
The type of data that transformers are designed to handle, i.e.\@ sequence data, makes them suitable not just for natural language problems but also for time-series data and signal processing. In the NLP setting, the output could be a translated text in a different language (multi-output) or a sentiment analysis prediction (single output), for example. The language input has a sequence-like structure due to the grammar and context of the sentence. Each word in a sentence can be seen as analogous to a time period in our data. In natural language applications, each word is converted to a vector of integers based on spelling, meaning, and other language attributes. Similarly, for each time period, we have many features that represent the financial state of the firm. When applied to language tasks, transformers apply multi-headed attention to each sentence and learn the sentence's relationship to the output. Here we apply a similar process over time series (sequence) data to learn how to predict default probability. 
    
\begin{figure}[t!]
 \begin{subfigure}[b]{.4\linewidth}
\includegraphics[scale=0.4]{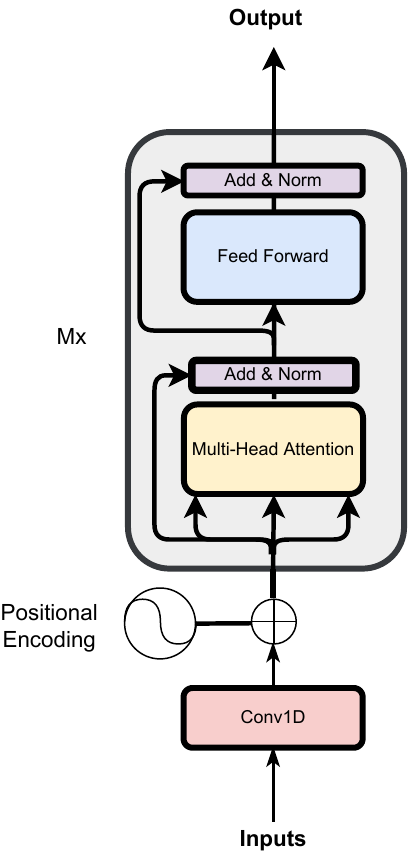} 
\setcounter{subfigure}{0}%
\caption{Transformer Encoder for Panel data (TEP)}\label{fig:CTE}
\end{subfigure}
 \begin{subfigure}[b]{.4\linewidth}
\includegraphics[ scale=0.5]{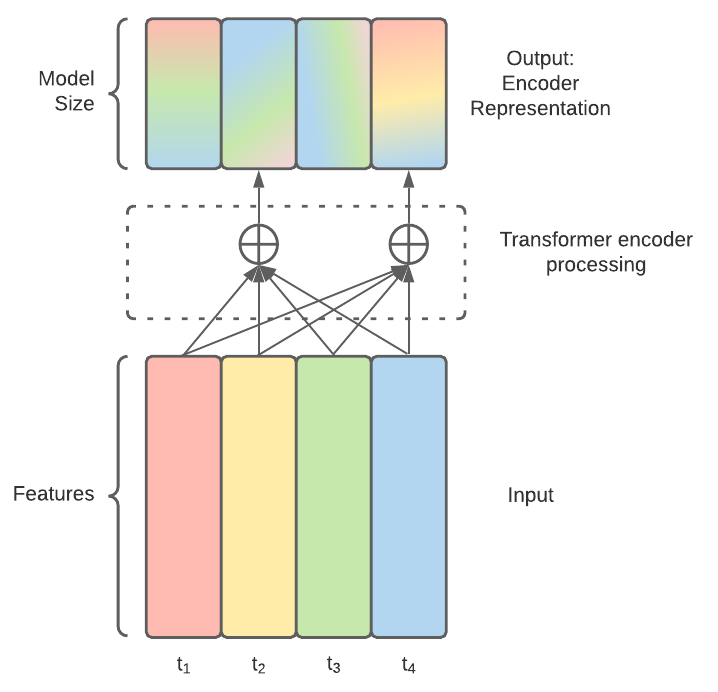}
\caption{ Encoder representation}\label{fig:EncoderRepresentation}
\end{subfigure}
\caption{Transformer Encoder architecture and representation}\label{fig:Transformer}
\end{figure}

Further advances are being made regarding the application of transformers to time series forecasting \citep{Li2019a,Wu2020a}. In this paper, we modify the original Transformer, by using only the encoder part to form a representation of the input data as shown in Figure~\ref{fig:Transformer}. The encoder representation is a more useful transformation of input data as the representation uses relationships across different times and also reshapes the data for it to be suitable for the task. In Figure~\ref{fig:EncoderRepresentation}, an example with four time periods of panel data and $f$ features is transformed to a representation at each time period with size $H$, which is the model size parameter in the transformer model.

For natural language tasks, using the encoder representation only has been quite successful as the BERT class of models originally developed by \citet{Devlin2018} shows. They use a similar architecture using the encoder part of a transformer and a few other deep learning layers specific to a task they are being trained for. As our problem is a multi-label classification task, we use the encoder output combined with a max-pooling layer and a dense layer. The max-pooling layer works as a filter leaving only those variables that maximise the signals found. The dense layer is a feed forward layer which modifies the encoder representations to suit our prediction target. This way, our transformer model encodes our set of time series into several feature vectors, which provide a detailed description of the company and its market context.  From the original transformer, we also modify the initial layer by replacing the embedding layer with a 1D convolutional layer as shown in Figure~\ref{fig:CTE}. This helps us in two ways. Firstly, unlike textual data that needs to be converted to numerical data accessible to the model, the time series data is already available in a numerical format. Secondly, transformer models have a fixed model size, which ensures a constant size flow of the input representation through each layer of the model. The initial convolutional layer modifies the time-series input to match the model size of the transformer model. This makes it possible to combine different data sources and model outputs, as we will show later. As the performance of the transformer proved sensitive only to the model size and number of layers, other aspects of the encoder are left unchanged.

As we suggested previously, the attention mechanism brings more interpretability to deep learning models. Self-attention in transformers is a more general form of attention, which is particularly applicable in our case, where we must extract the relationship between tokens in the same sequence. Compared with attention in RNNs, where the attention mechanism is generally global, it is placed between an encoder and a decoder to weight the encoder representation against the decoder output. In this paper, we are interested in the self-attention mechanism within the encoder. The architecture of RNN models is such that they build up a sequential data representation using the gated mechanism and are challenging to track over the entire period. In contrast, using the self-attention mechanism, transformers are designed to use the complete input sequence, making them more interpretable. The multi-headed attention mechanism allows different relations to be learnt from the same sequence, as shown in \ref{Appendix1}. Therefore, we believe transformers are suited to our task or any panel data classification tasks.

\subsection{Temporal Convolutional Network (TCN)}\label{subsec:TCN}

A temporal convolutional network is a generic architecture for sequence data \citep{Bai2018} which was found to give better results over benchmark models such as LSTMs and provides a good trade-off between model complexity and performance. TCNs can store a longer memory than LSTMs and are able to perform better when there are long-term persistencies in the data, as is the case for the financial performance of a company where losses or weak performance could persist over time.

\begin{figure}[t!]
 \begin{subfigure}[b]{.3\linewidth}
\includegraphics[width=\linewidth,height=0.3\textheight]{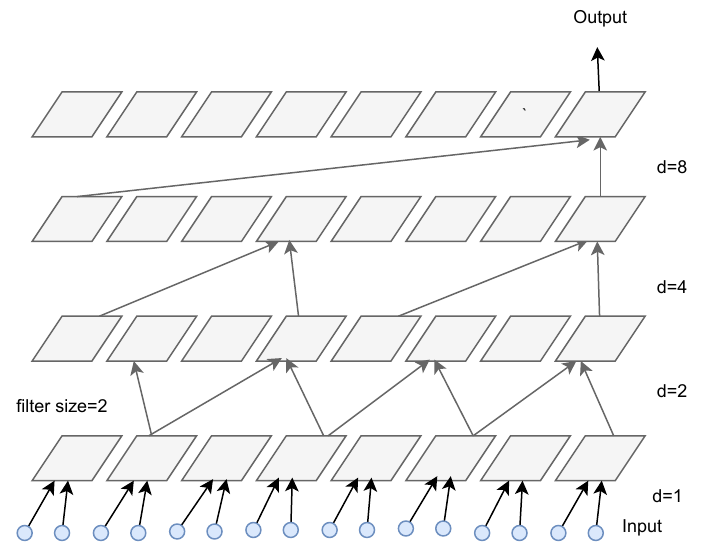}
\setcounter{subfigure}{0}%
\caption{Temporal Convolutional Network (TCN)}\label{fig:TCN}
\end{subfigure}
 \begin{subfigure}[b]{.6\linewidth}
\includegraphics[width=\linewidth]{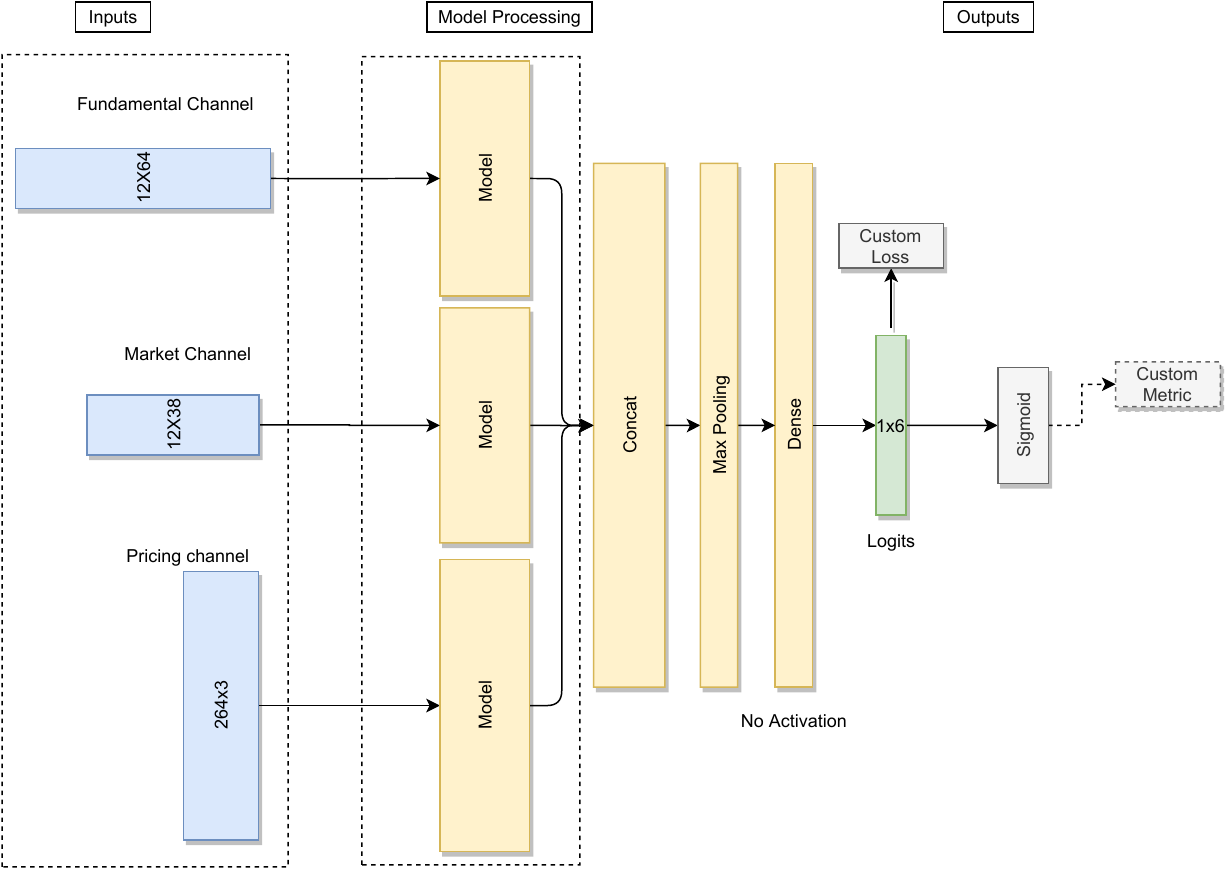}
 \caption{Multimodal deep learning model architecture}
  \label{fig:architecture}
\end{subfigure}
\caption{TCN model and multimodal architecture}\label{fig:TCNGenMod}
\end{figure}

TCNs build up a hierarchical memory over a sequence of data, as shown in Figure~\ref{fig:TCN}. Each row is made up of a number of residual blocks. Each residual block is made up of two dilated convolutional layers with weight normalisation and dropout. A dilated convolution is a convolution where the filter is applied over an area larger than its length by skipping input values with a certain step (d) \citep{Oord2016}. Initially at the input, the TCN looks at nearby relationships for data points and builds up a representation of the data. This process is repeated until we end up with one higher-level representation. Unlike transformers that focus on all data simultaneously, TCNs build their representation in a sequential manner. This makes them closer to image recognition models such as Convolutional Neural Networks (CNNs) but applicable to sequential data, including time-series data.

\subsection{Multimodal architecture} \label{sec:GeneralArchitecture}

One of our paper's contributions is that we develop a framework to add multiple data sources and combine them. To enable this, a multimodal approach is proposed, the architecture of which is given in Figure~\ref{fig:architecture}.
 
We could train this multimodal model in three different ways: train one data channel at a time; or train each model to its input but simultaneously; and finally, by using a differential training regime. In the first regime, we iterate over each data channel and train the relevant model. Once all the channels are processed, the multimodal model is ready for inference. Here the parameters of each model should converge closer to their global optima but we miss the interactions between data channels. In the second approach, we use all data channels so the models learn the interactions but could converge to local optima as all parameters of all the models are being learnt together. Thirdly, in the differential training approach, we combine the earlier two training methods. This kind of training approach is used in multi-task training setups \citep{Liu2019}, where a single model is trained on different tasks so the model improves its generalisation capabilities. Here we have one task but different models and different data. We initially train using the first approach and later train again so the models learn the interactions between different data channels. Instead of allowing all parameters to update, we only update the parameters of one or two channels to learn the effect of interactions. We chose to freeze the model where there is higher complexity of finding a relationship, i.e.\@ the pricing channel, followed by the market channel. For the fundamental channel, it is relatively easier to find a relation to the probability of default term structure. Also, updating only selected parts of the multimodal model improves the training time \citep{Lu2015}.
 
The arrows in Figure~\ref{fig:architecture} highlight the general data flow structure from inputs to outputs. The dotted lines around inputs or models mean they could be combined or run individually based on the analysis that we are looking to run. For example, if we are looking to use the fundamental channel data only, the other inputs will be disabled, and only one model will be used. The specific model that will be used for this data could be either TEP, TCN or LSTM, but the setup is easy to extend to other forms of data and models as well. The model layer in the architecture generates individual representations of its inputs which are processed further with three deep learning layers --- Concat, Max pooling and Dense layers. These are used in all three approaches of training. The Concat layer combines multiple representations, and the Max layer selects the representation with the highest values. The Dense layer uses these values to find interactions between these data sources. Among these only the Dense layer has parameters that need to be trained. The Dense layer outputs logits which are modified to predictions and are also used to calculate the loss for each batch of data.

\section{Model training and experiments}\label{sec:ExperimentalSettings}
This section describes the loss function, the Shapley method we used for the interpretation of the models, the hyper-parameter tuning strategy, the optimisation measures used during training, and the two testing strategies used.

\subsection{Loss function}
For model training, as we are dealing with an incremental multi-label classification problem, we need to define an appropriate loss metric. We chose to base ours on cross-entropy loss. With the last layer of the network outputting the logits ($ \hat y_{t}$) for our respective time horizons (i.e.\@ 3 months to 3 years), we enter each of those outputs into a sigmoid cross-entropy with logits function, defined as follows: 

  \begin{equation}
      L(y_t, \hat y_t) =
      -(y_{t}*\log(\text{sigmoid}(\hat y_{t}))+(1-y_{t})*\log(1-\text{sigmoid}(\hat y_{t})))
  \end{equation}
where $y_{t}$ denotes the true default outcome (0 or 1) for that outcome period, $t$. To obtain a loss value for the entire observation, we sum the loss values over all those time horizons. This loss function is different from the typical cross-entropy loss function for multi-class classification, as, instead of only one class having a positive outcome, we often observe multiple such outcomes depending on when the default occurred. Note that we do not have strict independence among the binary target vector variables as some combinations are not possible by definition. While we have not enforced that limitation in our current models (which could be done, e.g., by penalising the weights), it did not lead to incorrectly specified probabilities in our results.

\subsection{Shapley variable group importance}
We use Shapley values, a solution concept from game theory, to explain the relative importance of channels and the models' temporal dependence \citep{Nandlall2019}. Shapley values are calculated for the multimodal case by framing the problem in the form of a cooperative game. Playing a game is analogous to using the model to predict. Maximising the prediction metric is the objective, called the score function. 
 
The players in the game are the data channels defined earlier. If a channel is selected, it is denoted by 1, and 0 otherwise. For G channels, the universe of possible combinations is denoted by $U$ where $|U|=2^G$, and each combination is a profile $p_{i}$ where $i=1,2,..,2^G$. $|p_{i}|$ is the number of channels selected. When $|p_{i}|$ is 1, the profile is denoted by $e_{i}$, implying only one channel among the G channels is selected. The Shapley set ($Q_g$) of a channel $g$ is all the sets in $U$ in which channel $g$ is not selected ($g^{th}$ element is 0). 

The score function is a characteristic function taking only values between 0 and 1, a higher score indicating a more favourable outcome. While accuracy is often used as a score, in our setting, model performance is more often measured using the Area Under the ROC Curve (AUC). A higher AUC value suggests better ability to discriminate between defaults and non-default, but unlike accuracy, AUC does not take a zero value (but rather a value between 0.5 and 1), so to turn it into a valid score function, we need to rescale it into the so-called Gini coefficient (equal to 2*AUC-1) and use this as our score function.

The marginal contribution of the $i^{th}$ channel is dependent on the profile. For a profile $p_n$ where the $i^{th}$ channel is not included, the marginal contribution is the difference in score when the channel is added:

\begin{equation}
  m(p_n,e_i) = s(p_n+e_i)-s(p_n)
\end{equation}

The Shapley value for channel $i$, $S(i)$, is now defined as 

\begin{equation}
  S(i) =  \sum_{p_n \in Q_i} m(p_n,e_i)*(|p_n|)! (|G|-|p_n|-1)!/(|G|)! 
\end{equation}

In other words, $S(i)$ is the (weighted) average contribution of the $i^{th}$ channel to the game, weighing all possible combinations to which the channel can be added. A higher score implies a higher contribution of the channel's data towards the predictive power of the model. 

\subsection{Hyper-parameter tuning} \label{sec:hyper-parameterTuning}
We used a grid search to tune the hyper-parameters for each model, using a validation dataset covering 20\% of the total data. To speed up the search, we used parallel processing techniques. 

For logistic regression, we used the saga solver with L2 penalty, as it is easier to optimise than the L1 penalty but performed similarly in our experiments in terms of predictive performance.
    
The XGBoost model hyper-parameters were tuned with a grid search for the learning rate \{0.001, 0.01, 0.1\}, maximum depth \{2, 3, 4\}, number of estimators \{50, 100, 250, 500\} and alpha \{0.1,\ldots,0.9\}. These were found to be an appropriate choice of parameter ranges after trialling them on the different data channels. 
 
For the deep learning models, we found the batch size and number of epochs to be less important as we trained the models with early stopping, as explained later in section~\ref{sec:Othersettings}. The shallow neural network consisted of two hidden layers and one output layer. The first two layers were tuned over a different number of units in the range of \{50,100,150,200\} and \{10,20,30,40,50\}, respectively. In the LSTMs, we tuned the number of units, over the range \{16,32,64,96,128,150\}, the dropout rate \{0.1,0.2,0.3\} and the optimiser \{`adam',`sgd'\}. The TCN's hyper-parameters are different as it is a convolutional network-based model. There, we conducted a grid search on the number of filters \{16,32,64,128\}, kernel size \{1,3,6\}, the activation function \{`tanh',`relu'\} and dropout \{0.1,0.2,0.3\}. Finally, in the proposed TEP, the model size and number of layers are the key parameters that need to be determined. We tuned the model size ($M$) over \{6,12,18,24,36,48,54,72,84,96,102\} and based on validation data performance set it to 72.  The number of layers ($l$) was tuned over \{1,2,3,4,6,12\}. Once the layers and model size are fixed, $h$ or the number of heads is defined as $M/l$. All the other hyper-parameters in the model were unchanged from their defaults as the impact of further tuning them proved marginal.

To select the window size for the accounting input data, we experimented by training LSTM and TCN models with different window sizes of 4, 8 and 12. These represent lookback periods of 1, 2 and 3 years, respectively, as each year has four quarters of accounting data. Both models performed better with larger window sizes, implying that using a longer time span of financial data benefits deep learning, and that these methods have the capacity to process it. The same window size was applied across all models and combinations later on.  
         
As for the pricing channel, it has daily prices covering the previous two years, making the potential lookback period quite deep. We used a grid search for the appropriate window size for each model, trying window sizes of 3, 6, 9, 12 and 24 months. In the results section, we will report how the performance of each model changes with the choice of window size. 

\subsection {Model training and testing} \label{sec:Othersettings}
To prevent overfitting the data, we trained the models with early stopping, whereby training is stopped when the validation set loss metric no longer decreases. To avoid local minima, a patience setting of five (eight) was selected for the multimodal (single-channel) model setup, respectively. We apply more patience to single channel training as it is expected to take a larger number of epochs compared to the multimodal model whose parameters have already been tuned. This is especially true for the pricing channel where single-channel training ran for 30-40 epochs in our analyses, while the multimodal training only required 3 to 5 epochs.

All models were first assessed on an independent test set (20\% of the data), using AUC as the performance criterion. Furthermore, to assess the robustness of the model performance estimates, we also carried out a stratified 10-fold cross-validation procedure. This ensures the model is subjected to various changes in variable distributions and relationship or concept drift over time. Instead of the traditional procedure which would simply divide the training observations into 10 folds, we define the folds by assigning different companies to different folds; this ensures that observations linked to the same company appear in the same fold. We will report the average performance and variance across all folds.  

\section{Results and discussion} \label{sec:Results}

In this section, we present three sets of results.
We start by comparing models built using only one data channel at a time, to study their performance independently of the other channels, and to fix the pricing channel lookback window size. This also identifies the best set of models to apply in the multimodal architecture which uses all channels.
Secondly, the next subsection shows the results of our multimodal training and robustness checks for the different architectures. Thirdly, we compare the importance of the three channels using the Shapley approach.
 
\subsection{Model performance results} \label{ModelResults}

\subsubsection{Single-channel models} \label{subsec:Singlechannel}

\begin{figure}[b]%[b!]
\includegraphics[width=\linewidth]{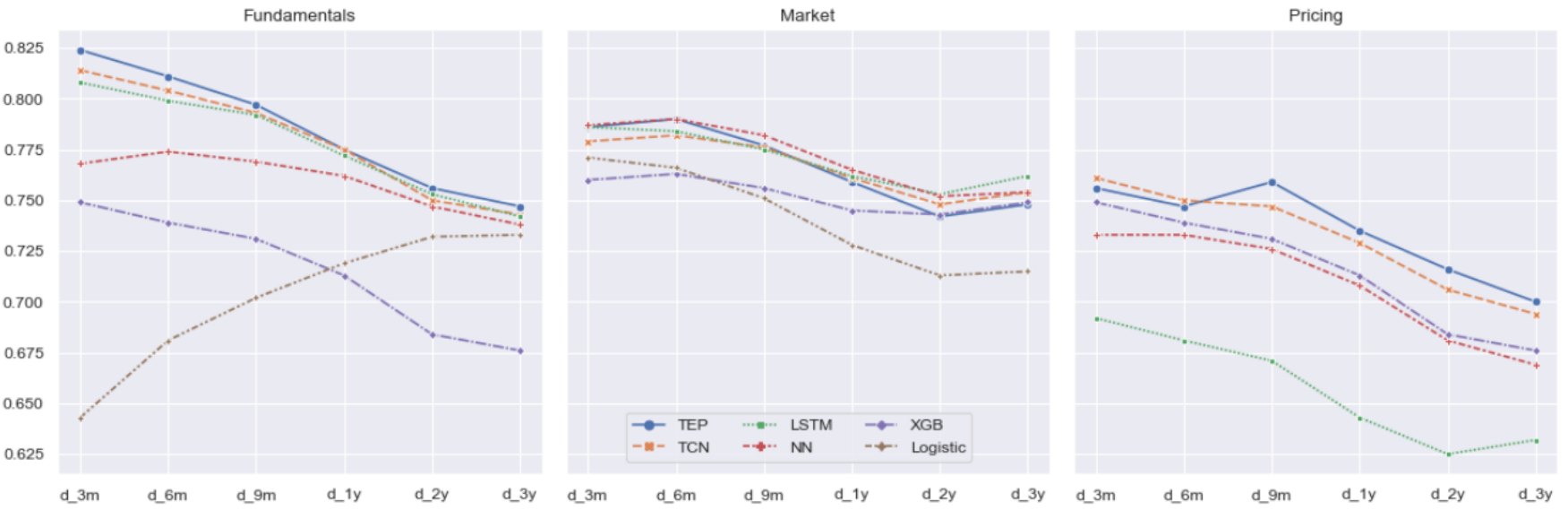}
\caption{AUC performance over different forecast horizons, for each data channel}\label{fig:allforecast}
\end{figure}

First, we consider each individual data channel as input and look to identify the best model for each channel. Table~\ref{tab:singlechannel} contains the AUC of each model, for each data channel, averaged over all the forecasting horizons. Figure~\ref{fig:allforecast} shows how the AUC performance for these models varies over the different horizons. This further tells us how the models perform over the short to medium term. To explain how the models for the pricing channel were affected by the choice of lookback window, Table~\ref{tab:WindowTuning} shows the performance for different window sizes. The best score for each window size is shown in bold and the best overall model and window size is underlined. Note that we dropped the logistic regression from this table as the AUC scores were not meaningful.

\begin{table}[t]
  \begin{tabular}{l|c|c|c}
    \toprule
    \textbf{Model} & Fundamental & Market & Pricing \\
    \midrule
    TEP   & \textbf{0.785} & 0.767 & \textbf{0.736} \\
    TCN   & 0.780 & 0.767 & 0.731 \\
    LSTM  & 0.777 & 0.770 & 0.657 \\
    NN    & 0.756 & \textbf{0.772} & 0.708 \\
    XGB   & 0.715 & 0.752 & 0.715 \\
    Logistic & 0.702 & 0.741 & 0.535 \\
    \bottomrule
    \end{tabular}%
    \caption{AUC model performance (averaged over all forecast horizons) for each data channel}
  \label{tab:singlechannel}%
\end{table}

Overall, the models for the fundamental data channel (see the left-most results column in Table~\ref{tab:singlechannel}) tend to perform better than those built for the other two channels, signalling the value of accounting data in predicting probability of default for mid-caps. Logistic regression (bottom row) was consistently the weakest performer among the models, especially with the pricing channel data. Other models were able to extract relationships that logistic regression could not, which suggests complex non-linear relationships between inputs and default risk that are not exploitable by general linear models.

The fundamental channel consists of quarterly accounting data, which is entirely firm-specific. The second column of Table \ref{tab:singlechannel} presents these results. With an average AUC of 0.785, the transformer (TEP) model gives the best performance for this data, but it is closely followed by the sequential deep learning models TCN and LSTM. Any of these models outperforms a shallow neural network model (NN), which has an AUC of 0.756. A potential explanation could lie in emerging complex structures that deeper models can better capture and local structures that are better represented in deep learning models (as opposed to global patterns that are equally captured by dense neural networks and logistic regression). Furthermore, in the leftmost graph of Figure \ref{fig:allforecast}, we can see that the deep learning models for the fundamental channel perform well over all forecast horizons. Interestingly, the performance gap between the different methods narrows the farther out the model has to predict default. This may point to a small number of features being more critical for longer-term default risk than any complex, temporal patterns.

The market data channel contains general market prices of several indices, as well as some company-specific data derived using them. For this channel, the NN model performs well, with an AUC of 0.772. The fact that, here, the sequential models do not outperform the NN suggests that there are not many temporal relations to learn in this data source. In other words, the deep learning models likely rely on other features useful for prediction. This result is intuitive as most of the temporal data in this channel is related to the general market environment, which does not impact a firm directly every day. Nonetheless, the data contains relevant information, as all models perform well in terms of AUC. These results show that, whereas complex networks are most useful when there is a large amount of firm-specific data from which complex relationships can be extracted (as was the case for the fundamental channel), embeddings derived from simpler networks are sufficient when considering varied data sources. 

Thirdly, the pricing channel contains just three features but has more frequent data points than the fundamental or market channels. The first question here was which look-back period or window size of past input data to select. To answer this, Table~\ref{tab:WindowTuning} lists the  AUC scores of each model (averaged again over all forecast horizons), for a series of alternative time windows.
\begin{table}[t]
    \centering
    \begin{tabular}{l|r|r|r|r|r}
    \toprule
   \multicolumn{1}{c}{\textbf{}}
      & \multicolumn{5}{c}{\textbf{Window size}} \\
    \midrule
    \multicolumn{1}{l|}{\textbf{Model}} & \multicolumn{1}{c|}{3m} & \multicolumn{1}{c|}{6m} & \multicolumn{1}{c|}{9m} & \multicolumn{1}{c|}{1y} & \multicolumn{1}{c}{2y} \\
    \midrule
    TEP & 0.698 & 0.710 & 0.711 & \textbf{0.716} & \underline{\textbf{0.736}} \\
    TCN   & 0.702 & \textbf{0.715} & \textbf{0.726} & 0.701 & 0.731 \\
    LSTM  & 0.588 & 0.654 & 0.626 & 0.570 & 0.657 \\
    NN    & \textbf{0.702} & 0.703 & 0.702 & 0.705 & 0.708 \\
    XGB   & 0.681 & 0.693 & 0.701 & 0.707 & 0.715 \\
    \bottomrule
    \end{tabular}%
    \caption{Pricing channel; AUC model performance for different lookback window sizes}
    \hfill
  \label{tab:WindowTuning}%
\end{table}
The results show that, as the pricing data's window size increases, the transformer model's AUC consistently improves, from 0.698 to 0.736. TCN, LSTM and, to a lesser extent, NN, also tend to improve with larger window sizes. Based on this, a two-year window was selected to produce all performance results. With an AUC of 0.736 and 0.731, respectively, it is the TEP and TCN models that are able to extract the most information from each such daily equity price window. As can be seen from the rightmost chart in Figure \ref{fig:allforecast}, the performance differences between the various models persist regardless of the prediction horizon. Notably the LSTM model could not extract long-term relationships similar to those that TCN and TEP were able to extract, and underperforms even when compared to other, simpler models. Overall, the performance for this channel remains lower compared to the other two channels, indicating that there is less predictive value in this type of data, or that the high level of noise in the pricing data cannot be filtered effectively by most methodologies.

As far as the term structure is concerned, Figure~\ref{fig:allforecast} confirms how, across all channels, prediction tends to become more difficult over longer time horizons. The only exception to this were the logistic regression models for the fundamental channel, which actually improve the farther out they are meant to predict.  
As we suggested earlier, the small set of accounting ratios on which those models rely may be more useful for longer term prediction. 
For the market channel, there is comparatively less of a drop-off in predictive power over a longer time horizon. This result shows the value of the general market environment for default prediction, especially over the medium to long term. Finally, the pricing channel results provide the clearest example of how model choice is crucial, as there are consistent performance differences between the different methods, regardless of the prediction horizon. 

Although our main focus is on predictive performance, not computation time, we note that the training times of tuned models are relatively short compared to the hyper-parameter tuning of the models. The models for the fundamental and markets data channels converged in 8 to 10 epochs (training time of approximately 20 minutes on two V100 GPUs), whereas the pricing data took around 30 to 40 epochs (training time of 40 minutes on two V100 GPUs). Again, as there is much more noise in daily pricing, it needs a more complex model to find a useful signal for default prediction, thus increasing training times.  

Next we look to integrate all sources of data into the modelling. However, combining high-frequency pricing data with low-frequency accounting data is not straightforward. Directly combining such data would require resizing the input matrices (e.g.\@ by turning quarterly data into daily values). Instead, deep learning provides several alternatives for building multi-channel models, as discussed next for the best performing model types identified thus far, i.e.\@ the TEP model and the TCN model.

\subsubsection{Multi-channel, all data} \label{subsec:Multichannel}
The multi-channel model is designed to use data from all three channels, using the architecture proposed in Figure~\ref{fig:architecture}. As this allows the three sets of inputs to be fed to the multimodal model separately, they can have different dimensions. We consider the different training approaches described in Section~\ref{sec:GeneralArchitecture}. In the previous section, we have already presented the results for the first training regime that evaluates individual data channels. Here, we first discuss how the models perform when we combine all data into one single input (see Table \ref{tab:rtable8}). Later, we present the results of the remaining training regimes.

Table \ref{tab:rtable8} shows that, when we merge all data channels, XGBoost performs best. We merge by converting the daily pricing channel into a quarterly channel, thereby dropping a large number of data points. This confirms some of the results in the literature where XGBoost performed well in classification tasks, including default prediction. The advantage of using a more sophisticated approach such as deep learning models, is that they would allow us to use separate channels, in their original form and thus reach an even higher accuracy. This advantage is not easily apparent if using data in a traditional way (as we do in Table \ref{tab:rtable8}). Instead, it requires providing the data in a manner that allows the feature engineering capabilities of these models to extract patterns that are not easily reproducible, as we show next.   

\begin{table}[t]%[!ht]
  \centering
  \caption{AUC model performance with all data sources combined}
      \begin{tabular}{l|c}
    \toprule
   \multicolumn{2}{c}{\textit{Input: All Quarterly data }}
   \\
   \midrule
    \textbf{Model} & \textbf{AUC} \\
    \midrule
    TEP   & 0.801 \\
    TCN   & 0.780 \\
    LSTM  & 0.800 \\
    NN    & 0.793 \\
    XGB   & \textbf{0.808} \\
    Logistic & 0.724 \\
    \bottomrule
    \end{tabular}%
  \label{tab:rtable8}%
\end{table}%

The results for the remaining two training regimes are presented in Table~\ref{tab:rtable6}, with the first column describing the training strategy applied to the data channels and the next column identifying the training regime, as described in Section \ref{sec:GeneralArchitecture}, to which it corresponds. The third column shows the average performance (over the different time horizons) of the multimodal architecture, using either TCN or TEP models. We chose to compare the TEP results with TCN, to understand whether the TEP model can outperform what was the next best model in the individual channel setup.
\begin{table}[!ht]
  \centering
  \caption{AUC performance of multi-channel TCN \& TEP models, for different training methods}
  \scalebox{.9}{
    \begin{tabular}{l|c|c|c|c|c|c|c|c}
    \toprule
    \multicolumn{2}{l}{\textit{Input: Quarterly and daily data channnels}} & \multicolumn{6}{c}{\textbf{AUC}}          &  \\
    \midrule
    \multicolumn{1}{l|}{\textbf{Method}} & \textbf{Regime} & Average & d\_3m & d\_6m & d\_9m & d\_1y & d\_2y & d\_3y \\
    \midrule
    \multicolumn{1}{l}{\textbf{TCN}} & \multicolumn{1}{r}{} & \multicolumn{1}{r}{} & \multicolumn{1}{r}{} & \multicolumn{1}{r}{} & \multicolumn{1}{r}{} & \multicolumn{1}{r}{} & \multicolumn{1}{r}{} &  \\
    \midrule
    Training together & 2     & 0.812 & 0.829 & 0.821 & 0.813 & 0.808 & 0.802 & 0.799 \\
    Pricing channel freeze & 3     & 0.817 & 0.828 & 0.833 & 0.822 & 0.812 & 0.805 & 0.802 \\
    Market and pricing channel freeze & 3     & 0.821 & 0.839 & 0.814 & 0.820 & 0.814 & 0.826 & 0.812 \\
    \midrule
    \multicolumn{1}{l}{\textbf{TEP}} & \multicolumn{1}{c}{} & \multicolumn{1}{l}{} & \multicolumn{1}{l}{} & \multicolumn{1}{r}{} & \multicolumn{1}{r}{} & \multicolumn{1}{r}{} & \multicolumn{1}{r}{} &  \\
    \midrule
    Training together & 2     & 0.835 & 0.858 & 0.848 & 0.843 & 0.832 & 0.820 & 0.812 \\
    Pricing channel freeze & 3     & 0.841 & 0.860 & 0.852 & 0.846 & 0.841 & 0.824 & 0.822 \\
    Market and pricing channel freeze & 3     & \textbf{0.847} & 0.867 & 0.860 & 0.850 & 0.847 & 0.833 & 0.824 \\
    \bottomrule
    \end{tabular}%
    }
  \label{tab:rtable6}%
\end{table}%

The prediction models clearly benefit from including all three channels in full, as the AUC values in Table~\ref{tab:rtable6} are larger than for all previously trained models. This shows how long time frames, varied data, and a complex data flow can lead to better results. The first table row tells us that the second training regime, for the TCN model, results in an AUC of 0.812. With this regime, we train all models together, so there is simultaneous updating of all model parameters. The third training regime further improves the model, to an AUC of 0.821 (see third data row), while for the TEP-based multimodal model, this regime produces an AUC as high as 0.847 (see bottom row) --- the best overall performance. With this approach, we train each model separately, thus enabling better optima to be found. 
We do so sequentially, first training the pricing channel as it performed the weakest among the three channels. Then we freeze the pricing channel model and train with both the pricing and market channel, allowing the interactions between both channels to be learnt. 
The same procedure is repeated later with fundamental channel data, allowing the fundamental channel to be trainable while we freeze the other two channels. We did not consider the direct interaction between fundamental and pricing channels here, as there was little improvement, but we tested all combinations as part of the Shapley interpretability experiment presented in Section~\ref{subsec:shapley}. Also, the architecture still allows for the interactions among all channels to be learnt through the final set of layers that combine these representations. Interestingly, as can be seen from right-hand side of the table, AUC improved with this regime over the (harder to predict) longer term as well, showing that long-term signals are being learnt.  This suggests that the differential training approach is better at handling the structural differences between the three input channels and consistently learns to derive relationships to predict over the complete term structure.

From Table \ref{tab:rtable6}, we also see that TEP consistently outperforms TCN, in all time horizons. It is particularly able to learn relationships for short-term default prediction, much more so than the TCN model. Long-term prediction is where the TCN model performance is relatively closer to that of the TEP model. 

Hence, an  avenue for further research could be to use different models for data channels based on the complexity of the data. Those could include applying TCN for the longer term, as it is less computationally intensive than TEP, or customised loss functions with different weights for the respective horizons. 

\subsection{Robustness check}
To test the robustness of our findings, we performed a 10-fold stratified cross-validation check for the multi-channel model, training sequentially using the previously described strategy, and assigning firms to different folds as described in  Section~\ref{sec:Othersettings}. 
\begin{table}[b]%[!ht]
  \centering
  \caption{Stratified k-fold cross validation: mean AUC (standard deviation)}
    \resizebox{\textwidth}{!}{%
    \begin{tabular}{l|l|l|l|l|l|l}
    \toprule
    \multicolumn{2}{c}{\textit{Stratified 10-fold cross validation}} & \multicolumn{5}{c}{\textbf{AUC}} \\
    \midrule
    \multicolumn{1}{l|}{Average} & \multicolumn{1}{l|}{d\_3m} & \multicolumn{1}{l|}{d\_6m} & \multicolumn{1}{l|}{d\_9m} & \multicolumn{1}{l|}{d\_1y} & \multicolumn{1}{l|}{d\_2y} & d\_3y \\
    \midrule
    0.869 (0.011) & 0.881 (0.025) & 0.884 (0.016) & 0.880 (0.013) & 0.871 (0.010) & 0.854 (0.008) & 0.846 (0.010) \\
    \bottomrule

    \end{tabular}%
    }
  \label{tab:rtable7}%
\end{table}%
Table~\ref{tab:rtable7} confirms that the proposed multimodal architecture produces excellent and stable default predictions, regardless of the time horizon. These results support the idea that the learning is able to detect true patterns as opposed to noise, as it successfully generalises to previously unobserved companies. Furthermore, the deep learning model can efficiently combine multiple information channels with limited preprocessing, even in the presence of significant noise.

\subsection{Interpretability of the architecture} \label{subsec:shapley}
Although the TEP was shown to produce accurate predictions, one challenge lies in providing a suitable interpretation of the factors that led to these predictions. To better understand the transformer model, one option is to interpret its multi-head attention weights. This analysis is provided in \ref{Appendix1} rather than the main paper, as such visual interpretations have been well studied. In this section, we instead chose to discuss the insights gained from the Shapley approach outlined earlier. Using this method, we can establish each data channel's relative importance and see how each combination of inputs has affected the AUC score.

\begin{figure}[!ht]
\centering

\begin{subfigure}[b]{.49\linewidth}
\includegraphics[width=\linewidth]{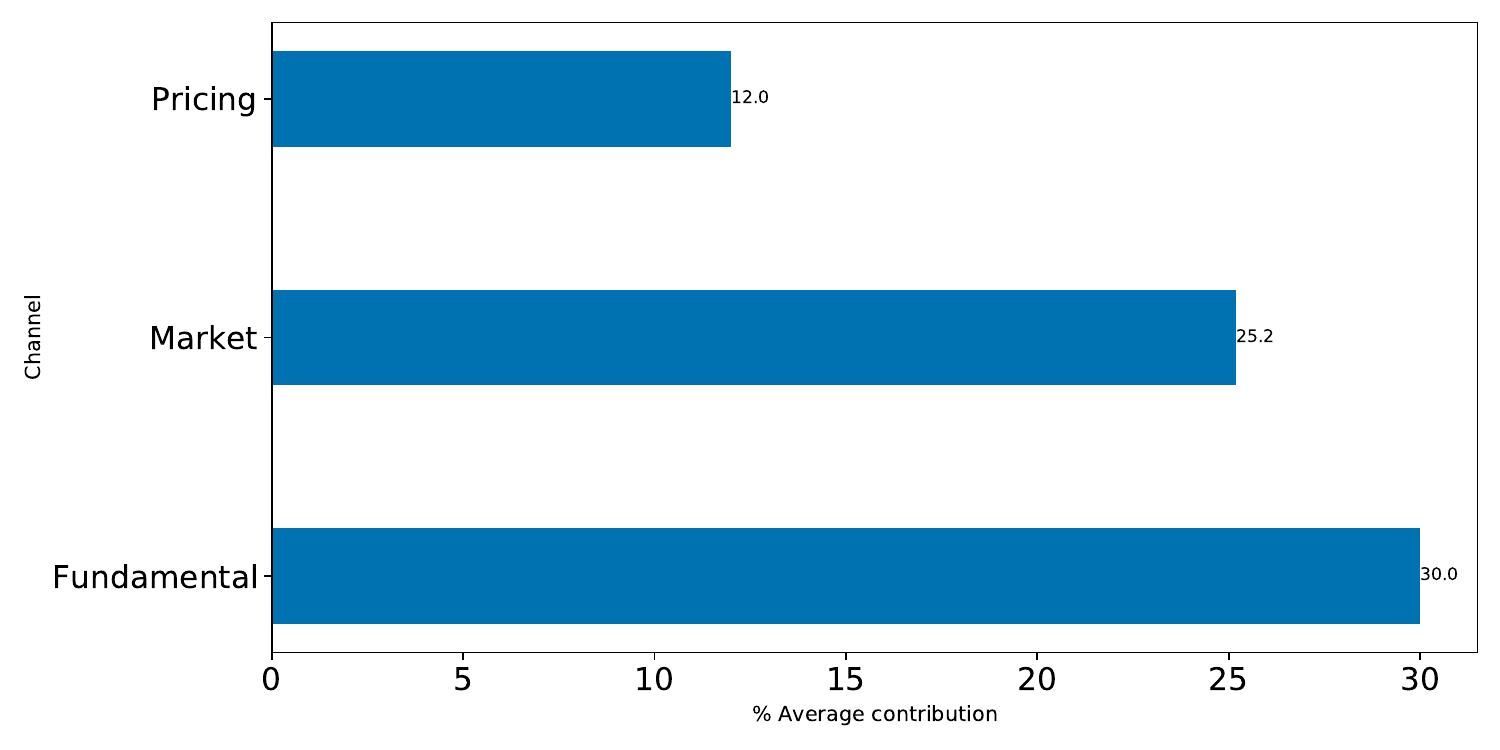}
\setcounter{subfigure}{0}%
\caption{Average contribution of each channel}\label{fig:avg}
\end{subfigure}
\begin{subfigure}[b]{.49\linewidth}
\includegraphics[width=\linewidth]{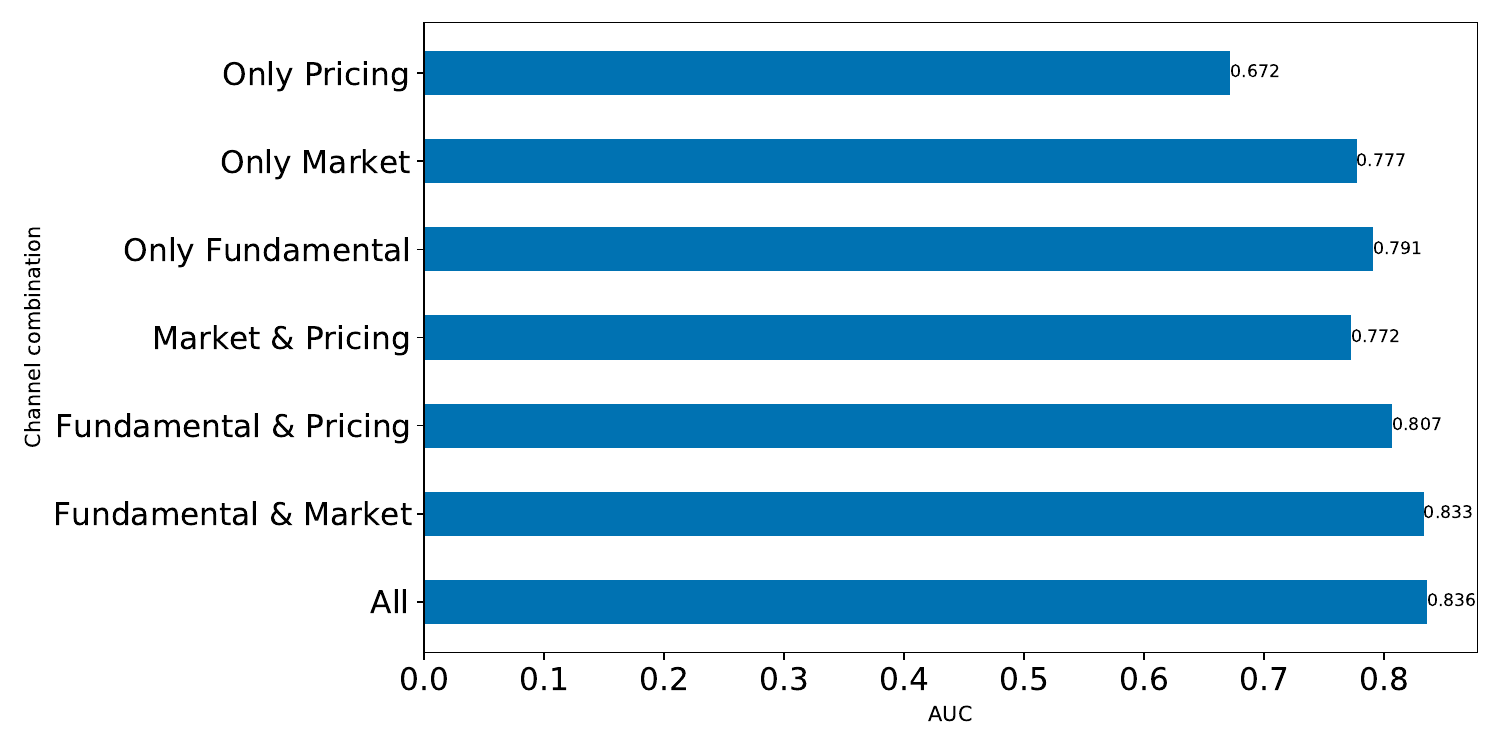}
\caption{AUC evolution over combination of channels}\label{fig:aucevo}
\end{subfigure}
\begin{subfigure}[b]{.92\linewidth}
\includegraphics[width=\linewidth]{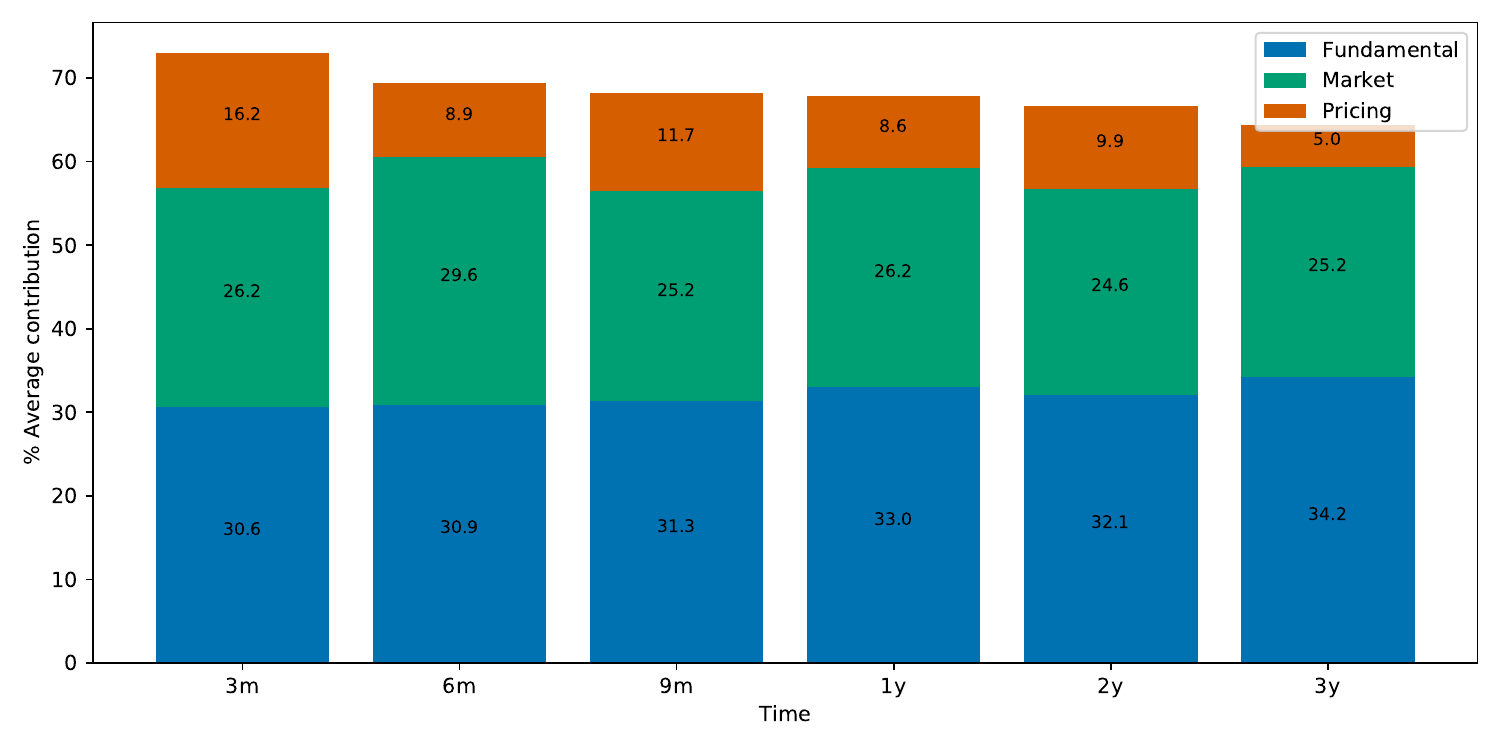}
\setcounter{subfigure}{2}%
\caption{Channel importance over time}\label{fig:imptime}
\end{subfigure}

\caption{Shapley analysis of channel importance}
\label{fig:shapley}
\end{figure}

As seen from Figure~\ref{fig:avg}, the fundamental channel has the highest relative contribution. On average, the inclusion of fundamental data into the model improves the model's AUC metric by 30 percentage points.  Figure~\ref{fig:aucevo} reports the AUC values for different combinations of data. For example, using only the fundamental channel, we achieve an AUC of 0.791. Adding the pricing channel slightly improves the performance to 0.807, while the market channel improves the AUC metric by 5.3\%, to 0.833. From both of these, it is clear that the market channel makes a larger contribution than the pricing channel.

To examine the impact of each channel more closely, we observe how the relative importance of channels varies over each prediction horizon in Figure~\ref{fig:imptime}. To predict default in the short term, the pricing channel plays a role with a contribution of 16.2\%, which decreases to just 5\% for the three-year horizon. From this, we infer that the pricing channel provides some signal in the short term. In contrast, over the medium term, fundamentals and the general market environment play a larger role in determining the probability of default. This follows intuition and somewhat aligns with the weak market efficiency hypothesis: prices reflect the market's current belief, taking into account short-term fluctuations, but true long-term estimation ignores these blips caused by events that may prove meaningless in hindsight.

Another temporal factor is which past time periods of input data contribute most to the model performance. To assess this, we divide the variables into two groups. Specifically, we group each twelve quarters of input data according to whether they belong to the most recent year, or the two years prior to that. Each channel is then evaluated on the test data to determine their relative importance. 

\begin{table}[!ht]
  \centering
  \caption{Shapley contribution of each channel over time (\%).}
    \begin{tabular}{l|c|c}
    \toprule
    \multicolumn{1}{r}{} & \multicolumn{1}{l}{\textbf{Shapley values}} &  \\
    \midrule
    \multicolumn{1}{l|}{\textbf{Channel}} & \multicolumn{1}{c|}{Past year} & \multicolumn{1}{c}{Previous 2 years} \\
    \midrule
    Fundamental & 52.3  & 12.4 \\
    Market & 35.1  & 20.0 \\
    Pricing & 38.4  & 9.3 \\
    \bottomrule
    \end{tabular}%
  \label{tab:PresentPast}%
\end{table}%

The results in Table~\ref{tab:PresentPast} highlight the importance of the latest time period, across all channels. In the fundamental channel, over 52\% of the performance comes from the financial performance reported in the most recent year. The time periods before that contribute positively to the model's predictions as well, but to a lesser extent (12.4\%). For the market channel, considering a longer time period becomes more important as, here, 20\% of the contribution comes from prior data. This could be expected as it takes some time for uncertainty in the macroeconomic environment to impact firms. In the pricing channel, the most recent data is contributing most towards performance. This implies that recent equity price trends are more informative in predicting (short-term) default rates. 

In summary, the fundamental data contributes the most to predict overall default, while the extensive amount of historical market data and historical price data further complement it, for example, to improve short-term default predictions. 

\section{Conclusion} \label{Conclusion}

The paper has shown that deep learning techniques, when carefully engineered, can predict complete term structures for credit risk, going beyond the one-year predictions hitherto common in the area. Seeing they outperform other methods on real-life mid-cap data, we find that the greater complexity of these models does increase predictive power. To achieve these performance gains, new strategies were needed. Specifically, we put forward a combined multimodal architecture, as this proved to be better than a single large model. This architecture simultaneously leverages market data alongside equity prices and the companies' fundamental data. It also gave us the flexibility to treat each data source differently and take advantage of selective learning mechanisms.

Suitable learning methods had to be devised for the problem at hand. For example, we used a loss metric for training purposes that is relevant to the incremental multi-label classification problem. We suggested this loss function could be further tweaked with a more intricate weighing mechanism for different horizons. Also, an efficient setup proved important as several models with different data combinations and different hyper-parameter choices had to be tuned. 

While the training strategy and suitable loss function can be applied in conjunction with any deep learning model, such as TCN or LSTM, we deployed a transformer-based model (TEP) and showed how to apply it to handle time-series-like structured data. The superior performance of this model over our data shows its promise in handling complex non-linear relationships over long time frames. TEP was able to handle lower-frequency data with many related features alongside high-frequency data, whereas other models experienced significant drops in performance when faced with such different data structures. 

As for our contribution towards advancing financial analysis, our results show that deep learning models can be successfully applied to mid-cap company default prediction, integrating different sources of data in a manner that traditional approaches could not. Mid-caps are often companies for which data could be missing or not be as widely available. Their prices could be more volatile, and they have a higher default rate compared to large-cap companies. Nevertheless, we were also able to show that accounting data still are the largest contributor to predicting default. The results also showed that market data are the second largest contributor and that pricing data can provide valuable additional signals in the short-term provided that we develop a differential training approach to handle this source of information. 

An often heard criticism of deep learning models is that they lack interpretability. To counteract this, we applied a method to interpret them using Shapley values for groups of variables; this method differs from other common SHAP-like approaches that produce an individual variable ranking instead. Using this approach, we could infer that pricing information is of some, albeit limited, time-decaying usefulness in the model, while the market context is much more important. Furthermore, we are able to visually infer the differences between defaulted firms and non-defaulting firms from the activation heatmaps derived from the TEP model, which naturally arise from attention-based layers. Added to the performance gains observed for them, being able to interpret TEP models in this manner also makes them a highly attractive deep learning method for a variety of credit risk settings like mortgage or credit card default predictions, where large-scale panel data is readily available as well.
     
An interesting avenue for further research is to extend the multimodal learning architecture put forward in our paper, by incorporating additional data channels, such as data related to the company's management, news feed data documenting relevant events or media coverage, etc. Although previous research has suggested there is value in such unstructured (e.g.\@ textual) data, little work has been undertaken yet to combine these alternative data sources along with the rich structured data used in this paper for the purpose of better understanding mid-cap default risk.  

\section*{Acknowledgements}
This work was supported by the Economic and Social Research Council [grant number ES/P000673/1]. The last author acknowledges the support of the Natural Sciences and Engineering Research Council of Canada (NSERC) [Discovery Grant RGPIN-2020-07114]. This research was undertaken, in part, thanks to funding from the Canada Research Chairs program.

\begin{appendices}

\section{Interpreting attention heat maps}\label{Appendix1}
We present a detailed interpretation of the attention weights to visually understand what kind of relationships are being learnt for mid-cap default prediction by transformer models. This kind of interpretation has been previously used in the NLP domain for translation tasks. To illustrate this, we select the fundamental channel data only. This gives a direct interpretation of the relationship between the TEP output and the input.

\begin{figure}[!ht]
\centering
\begin{subfigure}[b]{.45\linewidth}
\includegraphics[width=\linewidth]{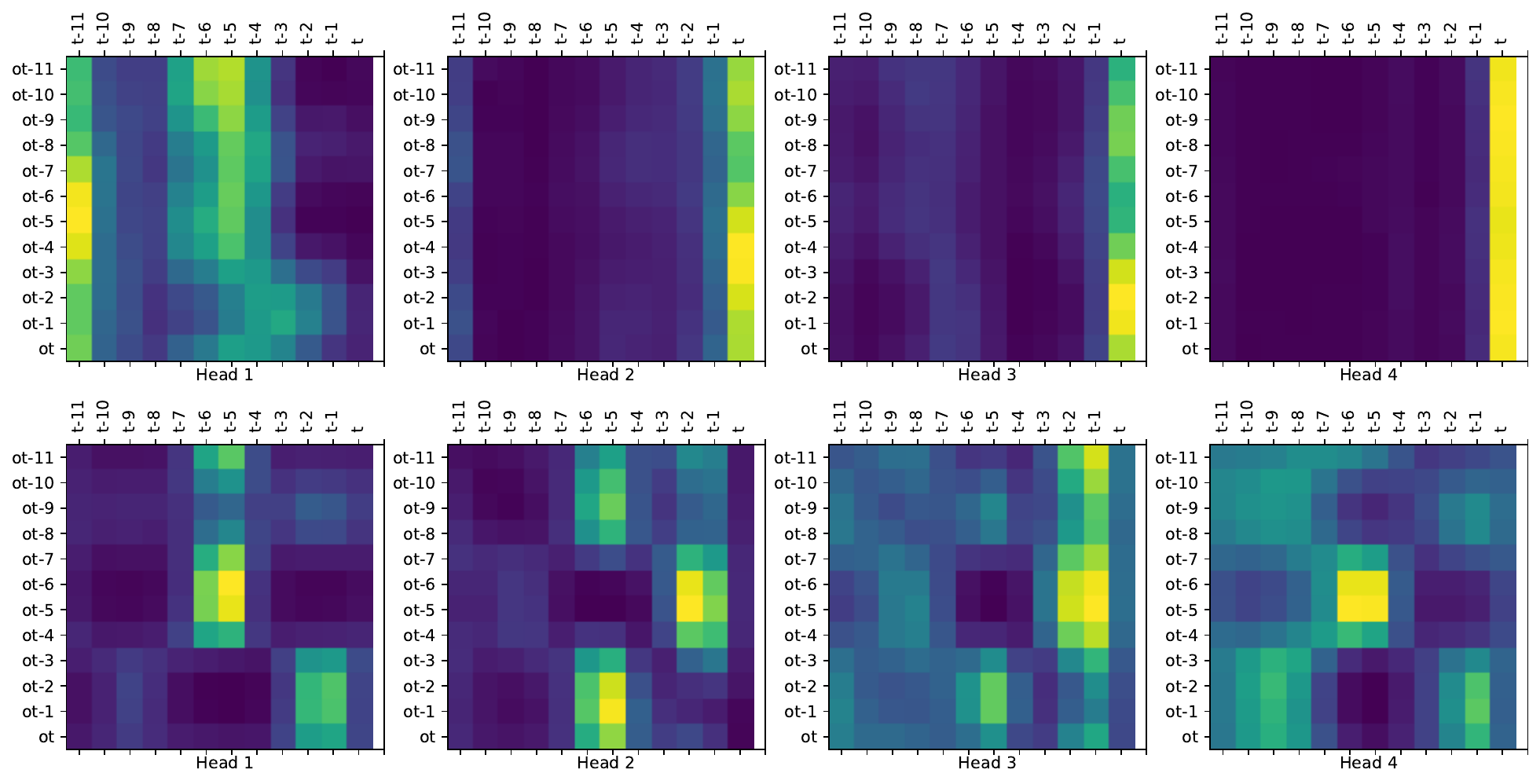}
\setcounter{subfigure}{0}%
\caption{Average weights for firms that default}\label{fig:defaultwgt}
\end{subfigure}
\begin{subfigure}[b]{.45\linewidth}
\includegraphics[width=\linewidth]{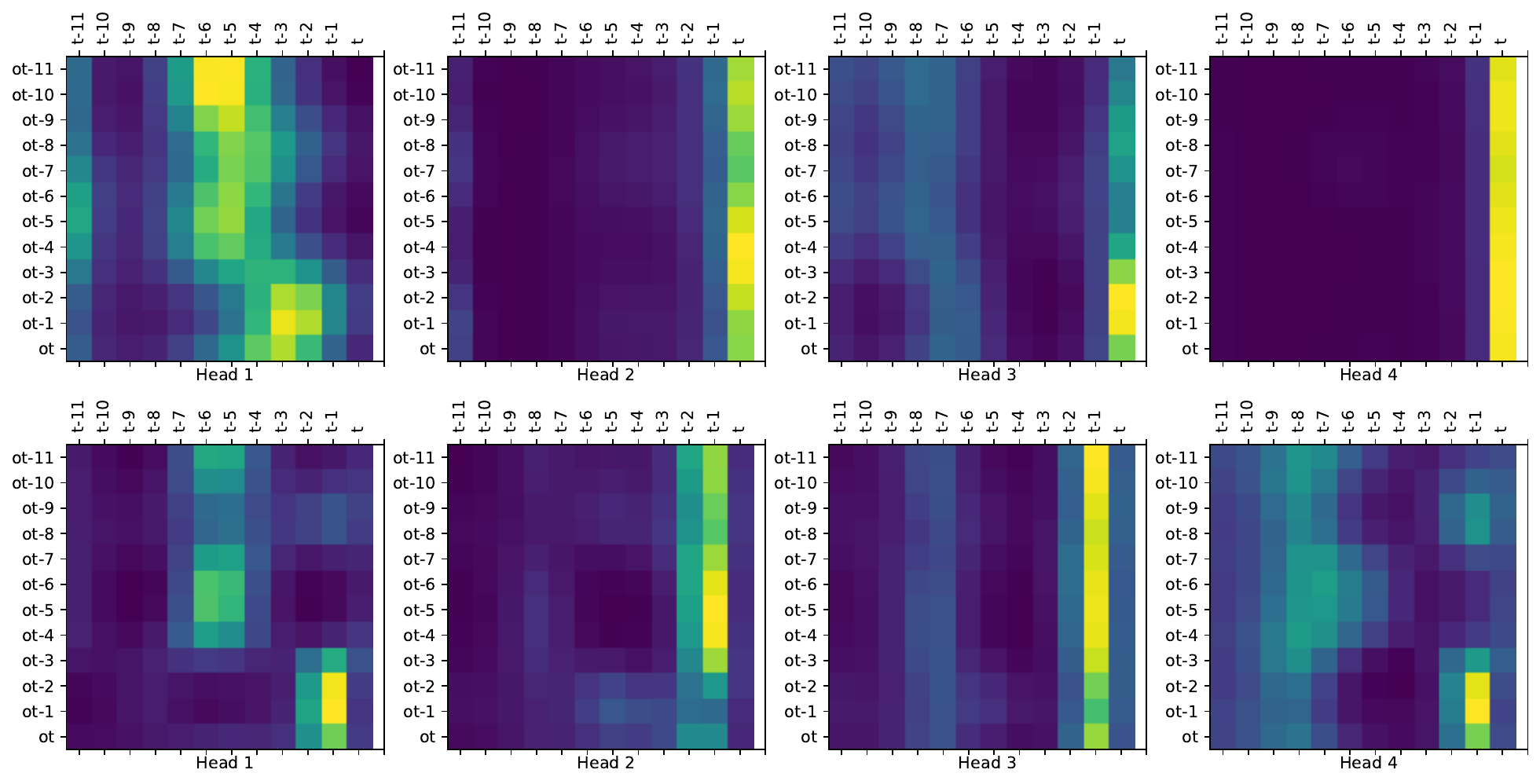}
\caption{Average weights of firms with no default} \label{fig:survivalwgt}
\end{subfigure}
\caption{Attention weights mapped to time periods, over default and non-defaults}
\label{fig:attentionweights}
\end{figure}

Each plot in Figure~\ref{fig:attentionweights} visualises the attention weights for one of the four heads (see the figure columns) in one of the two layers (rows) of the transformer model trained earlier. The horizontal axis in each plot divides the input data according to time quarter; the vertical axis is the output representation. This mapping thus shows which time period is given a higher weight by the head; the highest weights are shown in yellow, the lowest are in dark blue. To understand how the model distinguishes between default and non-default outcomes, we compare the average weights for firms that default (left panel) with those that do not (right panel). 

The first layer (top row) exhibits few differences between defaults and non-defaults. However, the second layer shows differences: the second head in the second layer, for defaulted firms, focuses on data from the $t-5$-th period and $t-2$-nd period, while the same head, for non-defaulted firms, looks at the $t-1$-th period. This can be interpreted as follows: if a firm has certain financial ratios in the last quarter of accounting data ($t-1$), it will be more likely to be classified as a non-default. If it does not satisfy this, the model looks at the previous financial year's data ($t-5$) to check for specific patterns to classify the firm as a default. This shows the model extracting complex temporal relationships. Other heads mainly use the present time period ($t-4$ to $t$) to extract relationships. 

In future work, those features highlighted in an attention heat map could be trialled as explanatory variables in a logistic regression, similarly to how one might employ other feature reduction techniques such as Principal Component Analysis. If this method improves the logistic regression model performance compared to using other techniques, this could be a value-add but it would be out of scope for this paper.  
\end{appendices}

\end{document}